\documentclass[10pt,aps,prb,twocolumn,superscriptaddress,nofootinbib,nobibnotes]{revtex4-2}

\usepackage[english]{babel}

\usepackage{amsmath}
\usepackage{amssymb}
\usepackage{mathtools}
\usepackage{braket}
\usepackage{bm}
\usepackage[pdftex]{graphicx}
\usepackage{xcolor}
\usepackage{placeins}

\usepackage{microtype}
\usepackage{soul}
\usepackage[toc,page]{appendix}
\usepackage{ulem}

\usepackage{hyperref}

\renewcommand{\vec}[1]{\bm{#1}}

\renewcommand{\d}{\mathrm{d}}

\def\ii{{\rm i}}

\begin{document}

\title{Hall resistance under a non-Hermitian linear potential}

\author{Kenan Uriostegui}
\email{rkuu@icf.unam.mx}
\affiliation{Instituto de Ciencias F\'isicas, Universidad Nacional
  Aut\'onoma de M\'exico, Cuernavaca, M\'exico}

\author{Santiago Galv\'an-y-Garc\'ia}
\email{santiagogyg@icf.unam.mx}
\affiliation{Instituto de Ciencias F\'isicas, Universidad Nacional
  Aut\'onoma de M\'exico, Cuernavaca, M\'exico}

\author{Thomas Stegmann}
\email{stegmann@icf.unam.mx}
\affiliation{Instituto de Ciencias F\'isicas, Universidad Nacional
  Aut\'onoma de M\'exico, Cuernavaca, M\'exico}
\date{\today}

\author{Jorge Lizarraga}
\email{jorge\_lizarraga@icf.unam.mx}
\affiliation{Instituto de Ciencias F\'isicas, Universidad Nacional
  Aut\'onoma de M\'exico, Cuernavaca, M\'exico}

\makeatletter
\def\Dated@name{}
\makeatother
\date{}
\begin{abstract}
We investigate how a non-Hermitian linear perturbation modifies Landau states and Hall transport, combining an exact continuum solution with numerical tight-binding calculations on a finite square lattice. The analytical solution yields energy branches with a complex linear Stark shift. A conserved non-Hermitian guiding-center operator identifies coherent states and determines the orbit center through its eigenvalue. The continuum model admits states with balanced gain and loss whose energies remain real at any perturbation strength. Finite-lattice calculations support the Landau-level structure and exhibit eigenvalue coalescences followed by the emergence of imaginary parts, consistent with exceptional points. Finally, transport calculations at a fixed injection energy show a shift of the Hall resistance plateaus moving them linearly from the von Klitzing constant.
\end{abstract}

\maketitle
\section{\label{sec:level1}Introduction}

An important feature of quantum mechanics is the Hermiticity property of the Hamiltonian operator, since it guarantees a real spectrum and the conservation of probability, and therefore admits a physical interpretation. 
However, Hermiticity is not the only condition that leads to a real Hamiltonian spectrum. As shown by Carl M. Bender and Stefan Boettcher \cite{bender1998real,bender2005introduction,bender2007making}, parity-time symmetry (${\cal PT}$-symmetry) can also yield a real spectrum of the system, even when the Hamiltonian is non-Hermitian. This discovery soon led to the development of time-dependent ${\cal PT}$-symmetric quantum mechanics, significantly expanding research in the field \cite{faria2007non,gong2013time,zhang2019time}. Beyond ${\cal PT}$-symmetry, non-Hermitian systems exhibit a rich symmetry structure. A complete classification of their symmetries was provided by Kawabata, Shiozaki, Ueda, and Sato \cite{kawabata2019symmetry}, who showed that 38 symmetry classes in total describe intrinsic non-Hermitian topological phases. Recently, three similarities shared among them have been pointed out \cite{Essential2024}. 

Experimentally, photonic systems with optical gain and loss 
have proven to be particularly important platforms for realizing non-Hermitian Hamiltonians \cite{bender2024pt,ashida2020non}. As noted in  \cite{feng2025non}, ``the capability to manipulate photonic eigenstates through optical gain and loss and non-reciprocal couplings provides a powerful toolbox for tailoring non-Hermitian Hamiltonians". Indeed, the paraxial equation of electromagnetic wave propagation has the same mathematical form as the Schr\"odinger equation with a complex potential \cite{feng2017non}, and a similar connection holds for the Helmholtz equation for optical cavities \cite{cao2015dielectric}. Hence, photonics has opened an avenue to explore non-Hermitian systems both experimentally and theoretically \cite{ozdemir2019parity,meng2024exceptional}. 

From a theoretical and numerical perspective, non-Hermitian Hamiltonians also become a powerful tool for describing non-equilibrium phenomena and transport processes \cite{ortega2020spectral,garcia2024atomically,sanchez2024edge,betancur2019electron,betancur2020phosphorene}. They play a central role in methods such as the non-equilibrium Green's function formalism \cite{stegmann2021brief} and give rise to genuinely new spectral features, including exceptional points (degeneracies where eigenvalues and eigenvectors coalesce) which have attracted interest due to their enhanced sensitivity to perturbations \cite{heiss2004exceptional,wiersig2020review,schomerus2024eigenvalue,bid2024uniform}. Whether in electronic, optical or acoustic systems non-Hermiticity has left its signature \cite{schindler2011experimental,chitsazi2017experimental,karthiga2016systems,feng2017non,bulgakov2008bound,eichelkraut2013mobility,alaeian2014non,ghaemi2021compatibility,longhi2015robust,zeuner2015observation,longhi2015non,christensen2016parity,shi2016accessing}. Recent experiments have demonstrated innovative approaches to designing non-Hermitian quantum Hall devices. For instance, a multiterminal quantum Hall ring featuring non-reciprocal hopping was investigated in Ref. \cite{ochkan2024non}, while an experimental setup for detecting the non-Hermitian quantum Hall effect in a Corbino ring was proposed in Ref. \cite{yan2024transport}. Furthermore, non-Hermitian dissipative $\mathcal{PT}$-symmetric lattices with controllable losses have been successfully realized \cite{eichelkraut2013mobility}.\\

Given its theoretical significance and the rapid experimental progress toward its realization, the impact of non-Hermiticity on well-known physical phenomena has attracted considerable attention. Among these, the Quantum Hall Effect (QHE) has been investigated in its integer~\cite{Ostahie2016, ochkan2024non}, fractional~\cite{Yoshida2019}, and anomalous (Chern-insulator)~\cite{yao2018non, Chen2018, Hirsbrunner2019, Kawabata2018, Bartlett2023, Liu2024, deOliveira2024, Ezawa2019} variants, and constitutes the focus of the present study. 
Integer quantum Hall effect (IQHE) is a phenomenon that occurs when a perpendicular magnetic field is applied to a two-dimensional material, forcing electrons in the bulk of the sample to form closed orbits, whereas electrons at the edge bounce \cite{klitzing1980new,von202040}. The Schr\"odinger equation of a particle in a magnetic field was solved by Landau, and the spectrum of the system is called Landau levels \cite{landau2013quantum}. The number of filled Landau levels determines the Hall conductivity, which has been proven to be quantized as $ie^{2}/h$ where $i$ is an integer and $h/e^{2}$ is the von Klitzing constant. This phenomenon is incredibly robust to perturbations \cite{jeckelmann1997high}, and it is attributed to the fact that the quantization of conductivity is topologically protected since the integer $i$ is a topological invariant, now known as Thouless-Kohmoto-Nightingale–den Nijs invariant, which is analogous to a Chern number \cite{thouless1982quantized}. This mathematical analogy inspired the development of topological band theory for Hermitian Hamiltonians \cite{hasan2010colloquium,qi2011topological,bernevig2013topological}. 
Attempts to generalize the band theory to non-Hermitian Hamiltonians have been made \cite{san2016majorana,lieu2018topological,zyuzin2018flat,yao2018non,zhou2018observation,shen2018topological,yokomizo2019non} and induced non-Hermitian perturbations on Weyl Hamiltonians have proven to preserve the topological charge and have shown that the addition of gain and loss could lead to a new class of topological phase transitions \cite{cerjan2018effects}. 

In contrast, the effects of non-Hermiticity on Hall conductivity have been studied by Philip, Hirsbrunner, and Gilbert, who proved that in the particular case of the non-Hermitian quantum anomalous Hall insulator, the Hall conductivity is no longer quantized despite being categorized as a non-Hermitian Chern insulator \cite{philip2018loss}. Hence, the effect of non-Hermiticity on the Hall conductivity is not trivial to predict.
Recent circuit experiments have demonstrated non-Hermitian control of Landau modes through artificial fields and non-reciprocal couplings~\cite{wang2026reshaping}. Spatially varying real and imaginary on-site potentials have also been implemented to observe continuum Landau modes~\cite{zhang2024continuum}. These advances motivate the study of complex linear potentials as a means of modifying the spectral and spatial properties of Landau states.

In this work, we investigate how a non-Hermitian linear perturbation modifies Landau states and Hall transport. We describe the system using a square-lattice tight-binding Hamiltonian under a uniform magnetic field and analyze its long-wavelength continuum approximation. The continuum problem admits an exact analytical solution, while finite-lattice calculations provide numerical support for the resulting Landau-level structure and gain-and-loss interpretation.

Our analytical results comprise explicit eigenfunctions and energy branches governed by a complex separation constant $\zeta$, together with the identification of a conserved non-Hermitian operator $\hat{\Pi}$. This operator reveals the coherent character of the states in the guiding-center degree of freedom and gives a spatial interpretation to $\zeta$, which determines the orbit center. At fixed $\zeta$, consecutive energy branches retain the Landau spacing and acquire a complex shift linear in the electric field \cite{stark1913beobachtung}. Real values of $\zeta$ select states with balanced net gain and loss and real energies for any perturbation strength within the continuum model. The finite-lattice spectrum additionally exhibits eigenvalue coalescences consistent with exceptional points.

We also derive an effective transverse response from the continuum current and relate it to the Hall response under an additive description of filled levels. Numerical transport calculations at a fixed injection energy show that the perturbation shifts the Hall resistance plateau, with an approximately linear dependence on $e{\cal E}$ over the investigated range. These results connect the analytical description of the states and currents with the perturbation dependence of a measurable transport quantity.

The remainder of the paper is organized as follows. Section~\ref{sec:square_lattice} presents the perturbed tight-binding model and its numerical spectral properties. Section~\ref{sec:solution} develops the continuum description and derives its eigenfunctions and energies. Section~\ref{sec:symmetry} identifies the conserved symmetry operator, explains the physical meaning of $\zeta$, and relates real eigenvalues to balanced gain and loss. Section~\ref{sec:current_for_all_n} examines the current distributions, the effective transverse response, and the numerical Hall resistance. Finally, Section~\ref{sec:conclusions} presents the conclusions.
\begin{figure}[t]
  \centering
  \includegraphics[scale=0.265]{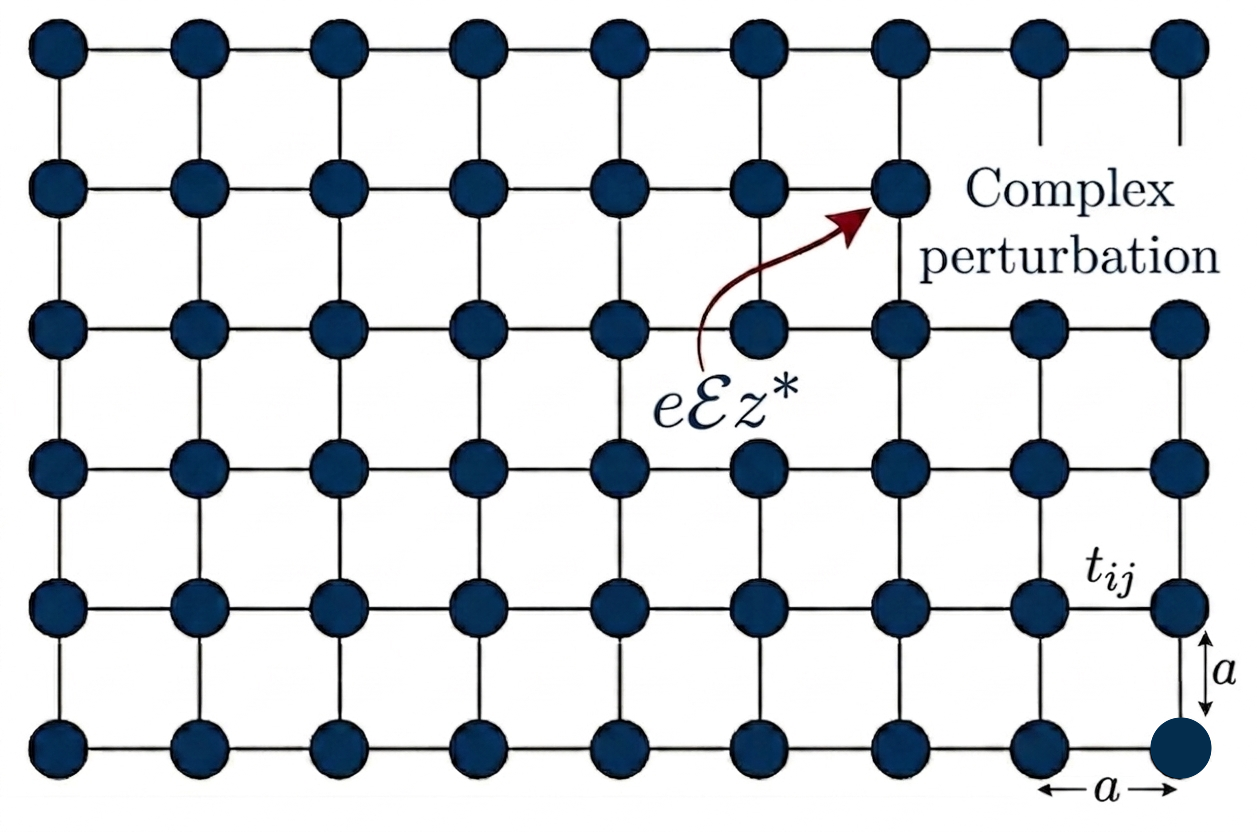}
  \caption{A square-lattice with lattice parameter $a$ representing the distance between atoms, $t_{ij}$ represents the hopping parameter between atoms. The Hermiticity of the lattice is lost due to the complex perturbation which is introduced in the onsite energy of each atom.}
  \label{fig:lattice}
\end{figure}
\begin{figure*}[ht]
  \centering
  \includegraphics[scale=0.28]{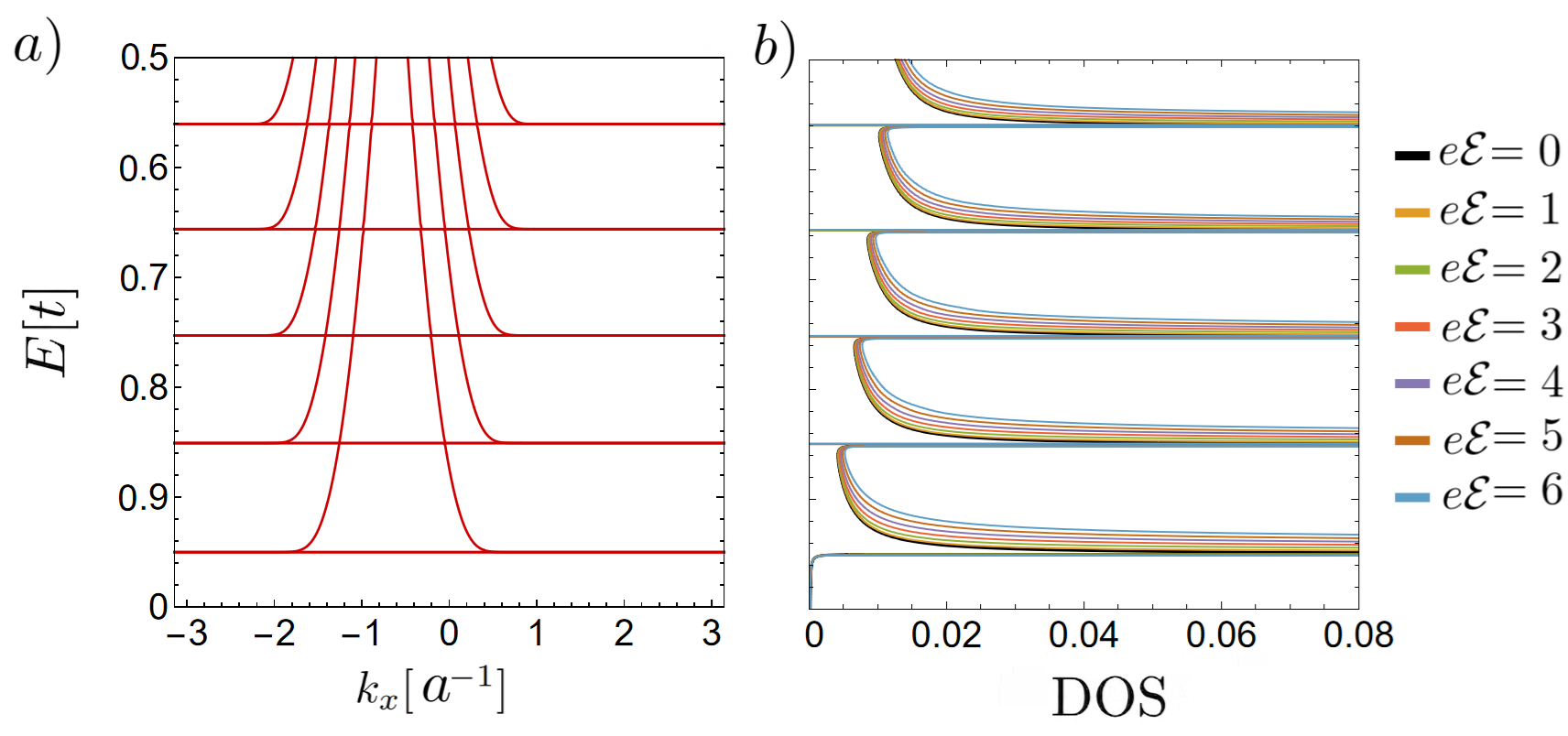}
  \caption{Panel $a)$, numerical band structure $E(k_x)$ exhibiting Landau levels for a square-lattice nanoribbon threaded by a uniform magnetic field without the perturbation. Periodic boundary conditions are imposed along the $x$ direction and open boundary conditions along the $y$ direction. Panel $b)$, density of states of a $350\times350$ square-lattice sample with open boundary conditions in both directions for different values of $e{\cal E}$ measure in units of $[\mu t/a]$. The peaks of the DOS coincide with the energies of the Landau levels, confirming the presence of Landau quantization in the finite sample. Energies are given in units of the hopping amplitude $t$.}
  \label{fig:nanoribbon_dos}
\end{figure*}
\section{\label{sec:square_lattice}Non-Hermitian square-lattice model }
The tight-binding model for a square-lattice can be written as
\begin{equation}\label{TBH}
\hat{H}_{0}=\sum_{i=1}^{N}\epsilon_{i}\ket{i}\bra{i}+\sum_{i\neq j}^{N}t_{ij}\ket{i}\bra{j}  
\end{equation}
where $\epsilon_{i}=4t$ is the onsite energy, $t_{ij}$ denotes the hopping amplitude between atoms $i$ and $j$. 
We retain only nearest-neighbor hopping. The vectors $\ket{i}$ form the standard orthonormal basis of the $N$-dimensional Hilbert space, where $N$ is the total number of lattice sites. 
The lattice sites are located at ${\bf r}=(x,y)=(na,ma)$, where $n,m$ are integers and $a$ is the nearest-neighbor distance.
The magnetic field can be introduced by the Peierls substitution on the hopping amplitudes
\begin{equation}\label{peierls}
    t_{ij}=-t\exp\left(-\ii\frac{e}{\hbar}\int\limits_{{\bf r}_{i}}^{{\bf r}_{j}}{\bf A}\cdot d{\bf l}\right)
\end{equation}
such that the hopping value is denoted as $t=1$ only between nearest neighbors and equal to zero other wise, $\vec{A}=(A_{x}, A_{y}, A_{z})$ is the magnetic vector potential such that the magnetic field is given by its curl, $\vec{B}=\nabla\times\vec{A}$, and the integration is performed from an atom in the position ${\bf r}_{i}$ to the position ${\bf r}_{j}$. All the quantities in this work will be expressed in terms of the hopping $t$ and the lattices distance $a$. The loss of Hermiticity is induced by introducing a complex perturbation 
\begin{equation}\label{eq:NHpert}
V(z^{*})=-e{\cal E}z^{*},
\end{equation}
where $z=x+\ii y$. This is equivalent to adding an onsite energy to the Hamiltonian \eqref{TBH} equal to this potential; see Fig. \ref{fig:lattice}. 
\begin{equation}\label{TBH2}
\hat{H}=\hat{H}_{0}+\sum_{i}V(z_{i}^{*})\ket{i}\bra{i}.
\end{equation}

The square-lattice spectrum develops a Landau-level structure near the lower band edge. Figure~\ref{fig:nanoribbon_dos} illustrates this structure through the nanoribbon band spectrum and the density of states (DOS) of a finite square sample.
We choose the Landau gauge $\mathbf{A} = (-By, 0, 0)$, which allows us to impose periodic boundary conditions along the $x$-direction and open boundary conditions along the $y$-direction. This gauge choice makes the crystal momentum $k_x$ a good quantum number, enabling a Fourier transformation of the Hamiltonian \eqref{TBH} along the $x$-axis:
\begin{equation}
    \ket{k_{x}}=\frac{1}{\sqrt{N_x}}\sum_{n_x}e^{ik_{x} n_x a}\ket{n_x},
\end{equation}
where $N_x$ is the number of sites along $x$. This transformation reduces the 2D problem to a set of 1D Hamiltonians parameterized by $k_x$. In this nanoribbon geometry, the magnetic field, chosen such that the magnetic field, divided by the magnetic flux quantum $\phi=eB/h$, is set to $\phi=0.008 [a^{-2}]$, induces edge states, which manifest as dispersive bands connecting the flat bulk Landau levels across the energy gaps, see Fig. \ref{fig:nanoribbon_dos} ($a$).

The position dependence of the perturbation in Eq.~\eqref{eq:NHpert} breaks translation invariance along the ribbon, so the perturbed Hamiltonian cannot be reduced to independent blocks labeled by $k_x$ \cite{yao2018edge,yokomizo2019non}. 
To examine how the non-Hermitian perturbation \eqref{eq:NHpert} affects the Landau-level structure identified in the unperturbed band calculation Fig.~\ref{fig:nanoribbon_dos} (a), we calculate the DOS of a finite $N_x\times N_y$ lattice with open boundary conditions for different values of $e\mathcal{E}$.
As shown in Fig.\ref{fig:nanoribbon_dos} ($b$), the singularities associated with the bulk Landau-levels remain distinguishable throughout the investigated range of the parameter $e{\cal E}$. Increasing the perturbation broadens these singularities without destroying the underlying Landau-level structure. These numerical results motivate us to investigate whether this structure persists in the continuum description.


\section{\label{sec:solution}Solution of the continuum model} 

It is known that the discrete Hamiltonian of a square lattice approximates the continuum Schrödinger Hamiltonian in the long-wavelength limit; see references \cite{isozaki2022continuum,chang1997multifractal,harper1955single}. Hence, the continuum Hamiltonian can be written as:
\begin{equation}\label{Heff}
    \frac{1}{2m}\left(\hat{\bf{P}}-e\vec{A}\right)^{2}\psi+V\psi=E\psi,
\end{equation}
where we defined the effective mass as $m = \hbar^2 / (2 a^2 t)$, from which the hopping energy parameter is related $t=\hbar^2/(2 m a^2)$, $E$ are the eigenvalues of the tight-binding model \eqref{TBH2}, consequently, all energies are expressed in units of $t$; $\psi=\psi(x,y)$ is the continuous wave function, and the linear momentum operator is defined as  $\hat{\bf{P}}=-\ii\hbar\nabla$. 

To analyze the continuum model, it is particularly useful to describe the same magnetic field using the symmetric gauge 
\begin{equation}
    \vec{A}(x,y,z)=\frac{B}{2}(-y,x,0),
\end{equation}
where $B$ is the magnetic field intensity. Hence, substituting the expressions for the potentials in the Hamiltonian (\ref{Heff}), expanding the squared terms, and introducing the cyclotron frequency $\omega_{c}=eB/m$, one can write the following expression
\begin{equation}\label{Heff_z}
    \hat{H}=\frac{1}{2m}\left(\hat{p}_{x}^{2}+\hat{p}_{y}^{2}-m\omega_{c}\hat{L}_{z}+\frac{m^{2}\omega_{c}^{2}}{4}(x^{2}+y^{2})\right)-e{\cal E}z^{*},
\end{equation}
where the angular momentum $\hat{L}_{z}=x\hat{p}_{y}-y\hat{p}_{x}$ was defined. To address this system, it is useful to begin by changing the coordinates to the complex plane, such that $z=x+\ii y$ and $z^{*}=x-\ii y$. Then the eigenvalue equation can be written as $\hat{H}\psi=E\psi$ where $\psi=\psi(z,z^{*})$, and can be simplified by using a Gaussian factor in the wave function
\begin{equation}\label{substitution}
  \psi(z,z^{*}) = e^{-\alpha zz^{*}}\phi(z,z^{*}),\quad 
  \alpha=\frac{m\omega_{c}}{4\hbar}=\frac{1}{4l_{B}^{2}},
\end{equation}
The quantity $l_{B}$ is known as the magnetic length and represents the characteristic radius of cyclotron orbits. 
The eigenvalue equation is simplified to the next equation
\begin{equation}\label{eq:eqSimplified}
    -\frac{\partial^{2}\phi}{\partial z^{*}\partial z}+2\alpha z^{*}\frac{\partial \phi}{\partial z^{*}}-\frac{m}{2\hbar^{2}}e{\cal E}z^{*}\phi=\left(\frac{mE}{2\hbar^{2}}-\alpha\right)\phi.
\end{equation}
The variables in this expression can be separated by proposing the function $\phi$ to be the product of two functions that depend only on their respective variables, that is, $\phi(z,z^{*})=f(z)g(z^{*})$, and the following equations are obtained 
\begin{equation}\label{ecuacion1}
    \frac{\partial f}{\partial z}=\zeta f,
\end{equation}
and
\begin{equation}\label{equation_g}
    (2\alpha z^{*}-\zeta)\frac{\partial g}{\partial z^{*}}-\frac{m}{2\hbar^{2}}e{\cal E}z^{*}g=\left(\frac{mE}{2\hbar^{2}}-\alpha\right)g,
\end{equation}
where $\zeta\in\mathbb{C}$ is a complex integration constant.
Both equations are solvable analytically, which gives the functions
\begin{equation}\label{eq:solTrivial}
    f(z)=e^{\zeta z},
\end{equation}
and
\begin{equation}
    g(z^{*})=\exp\left(\frac{m}{2\hbar^{2}}\frac{e{\cal E}}{(2\alpha)^{2}}(2\alpha z^{*}-\zeta)\right)(2\alpha z^{*}-\zeta)^{\rho},
\end{equation}
where $\rho$ was defined as
\begin{equation}\label{eq:rhoDefinition}
    \rho=\frac{m}{2\hbar^{2}}\frac{e{\cal E}}{(2\alpha)^{2}}\zeta+\frac{1}{2\alpha}\left(\frac{mE}{2\hbar^{2}}-\alpha\right).
\end{equation}
The eigenvalues of the system are obtainable by expressing the function $g(z^{*})$ as 
\begin{equation}\label{eq:partialSolution}
    g(z^{*})=\exp\left(\frac{m}{2\hbar^{2}}\frac{e{\cal E}}{(2\alpha)^{2}}(2\alpha z^{*}-\zeta)\right)\sum_{n=0}^{\infty}a_{n}(z^{*})^{n}.
\end{equation}
where $a_{n}\in\mathbb{C}$ are complex constants, substituting it in Eq.(\ref{equation_g}) and rearranging, one obtains the recurrence relation for $a_{n+1}$ in terms of its predecessor $a_{n}$
\begin{equation}\label{eq:recurencerRelation}
    a_{n+1}=-\frac{\frac{mE}{2\hbar^{2}}+\zeta\frac{m}{2\hbar^{2}}\frac{e{\cal E}}{2\alpha}-2\alpha n-\alpha}{\zeta(n+1)}a_{n}.
\end{equation}
For $\zeta\neq0$, a non terminating series satisfies
\begin{equation}
    \lim_{n\rightarrow\infty}\frac{a_{n+1}}{a_n}
    =\frac{2\alpha}{\zeta}.
\end{equation}
This gives a finite radius of convergence $|\zeta|/(2\alpha)$. To determine the admissible energies, we require the wave function to be single-valued and regular throughout the plane, including the point $2\alpha z^{*}-\zeta=0$. We assume that no singular point is excluded from the configuration space. The exponent in the explicit solution for $g$ must therefore satisfy
\begin{equation}
    \frac{m}{2\hbar^{2}}
    \frac{e\mathcal{E}}{(2\alpha)^{2}}\zeta
    +\frac{1}{2\alpha}
    \left(\frac{mE}{2\hbar^{2}}-\alpha\right)
    =n.
\end{equation}
For $\zeta\neq0$, this condition makes the series terminate at order $n$. For $\zeta=0$, the same condition follows directly from the explicit solution for $g$, without using the recurrence relation. The resulting energies are
\begin{equation}\label{eq:energies}
    E_n(\zeta)
    =\hbar\omega_c\left(n+\frac{1}{2}\right)
    -2\frac{\hbar}{m\omega_c}e\mathcal{E}\zeta.
\end{equation}
The first term reproduces the Landau energies \cite{landau2013quantum}, whereas the second introduces a complex shift linear in the electric field. Its real part represents a linear Stark shift \cite{stark1913beobachtung}. At fixed $\zeta$, consecutive branches remain separated by $\hbar\omega_c$.

The eigenfunctions in Cartesian coordinates are
\begin{widetext}
\begin{equation}\label{waave_func}
    \psi_{n}(x,y;\zeta)=A_{n}\exp\left(-\alpha (x^{2}+y^{2})+\zeta (x+\ii y)+\frac{m}{2\hbar^{2}}\frac{e{\cal E}}{(2\alpha)^{2}}(2\alpha(x-\ii y)-\zeta)\right)(2\alpha (x-\ii y)-\zeta)^{n}
\end{equation}
\end{widetext}
where $A_{n}$ is the normalization constant and has to be determined by means of the inner product. A biorthogonal formulation is commonly used to describe non-Hermitian systems, see Ref.~\cite{sim2025observables} for a discussion. However, in the continuum model with $e\mathcal{E}\neq0$ and $\alpha>0$, the formal adjoint solutions obtained by an analogous separation contain the growing factor $\exp(\alpha|z|^{2})$ and are not square-integrable. These separated solutions therefore do not provide a biorthogonal normalization in $L^{2}(\mathbb{R}^{2})$. We instead use the standard inner product to normalize the right eigenfunctions. In the unperturbed case $\mathcal{E}=0$, the Hamiltonian reduces to the self-adjoint Landau operator, for which normalizable adjoint eigenfunctions are available. Using the standard inner product yields the following normalization constant (see Appendix~\ref{app:normalization}): 
\begin{equation}
  A_{n} = \left[\pi n!(2\alpha)^{n-1}\exp\left(\frac{\gamma^{2}}{4}+\frac{|\zeta|^{2}}{2\alpha}\right)L_{n}\left(-\frac{\gamma^{2}}{4}\right)\right]^{-1/2}
\end{equation}
where $L_{n}$ are the Laguerre polynomials and 
\begin{equation}
  \gamma = \frac{m e \mathcal{E}}{(2\alpha)^{3/2}\hbar^{2}}. \label{eq:gamma}
\end{equation}


\section{\label{sec:symmetry}System symmetry}

To elucidate the role of the separation constant $\zeta$ in the solution of the system, we compute the first-order symmetry operator $\hat{\Pi}$ such that $\left[\hat{\Pi}, \hat{H}\right]=0$ (see Appendix \ref{app:Symmetries}). 
We find that the symmetry operator $\hat{\Pi}$ takes the form
\begin{equation}
    \hat{\Pi} = \ii \hat{p}_{x} + \hat{p}_{y} + 2\hbar \alpha (\hat{x}-\ii\hat{y}).
\end{equation}
By applying the operator to the state (\ref{waave_func}), one obtains the following eigenvalue equation
\begin{equation}
    \hat{\Pi}\psi_{n}(x,y;\zeta) = 2\hbar \zeta\,  \psi_{n}(x,y;\zeta).
    \label{eq:Pi-eigenvalue}
\end{equation}
Therefore, $\zeta$ is a continuous label associated with the eigenvalue of the conserved operator $\hat{\Pi}$. The relation $[\hat{\Pi},\hat{\Pi}^{\dagger}]=2\hbar^{2}/l_{B}^{2}$, together with Eq.~\eqref{eq:Pi-eigenvalue}, identifies the states $\psi_n$ as coherent states associated with the Heisenberg-Weyl algebra generated by $\hat{\Pi}$ and $\hat{\Pi}^{\dagger}$.

In contrast to $\hat{\Pi}$, the operator $\hat{\Pi}^{\dagger}$ does not commute with the Hamiltonian, $\left[\hat{\Pi}^{\dagger}, \hat{H}\right] = 2\hbar e  \mathcal{E}$. 
To understand the implications of this difference, the symmetry operator is placed in the context of the unperturbed Landau problem \cite{goerbig2009quantumhalleffects,tong2016quantumhalleffect}. By expressing the operator $\hat{\Pi}$ in complex coordinates, one obtains
\begin{equation}
    \hat{\Pi} = 2\hbar \left(\frac{\partial}{\partial z} + \alpha z^{\ast}\right) = \frac{\ii\sqrt{2}\hbar}{l_{B}}\hat{b},
    \label{eq:Pi-b}
\end{equation}
with 
\begin{equation}
    \hat{b} = -\ii\sqrt{2} \left(l_{B} \frac{\partial}{\partial z} + \frac{1}{4l_{B}}z^{\ast}\right). \label{eq:relacion-momentos}
\end{equation}
The operators $\hat{b}$ and $\hat{b}^{\dagger}$ generate a Heisenberg–Weyl algebra, $\left[\hat{b}, \hat{b}^{\dagger}\right] = 1$, and commute with the Hamiltonian
\begin{equation}
\hat{h} = \frac1{2m}\left(\mathbf{\hat{P}}-e\mathbf{A}\right)^2.
\end{equation}
They therefore constitute the symmetry algebra of the Landau problem, providing a discrete quantum number that labels the solutions of $\hat{h}$. Through these operators, the so-called guiding-center operators are defined,
\begin{equation}
    \hat{X}_{\hat{b}} = \frac{l_{B}}{\ii \sqrt{2}}\left(\hat{b}^{\dagger}-\hat{b}\right) \quad \text{and} \quad \hat{Y}_{\hat{b}} =  \frac{l_{B}}{\sqrt{2}}\left(\hat{b}^{\dagger}+\hat{b}\right)
\end{equation}
which specify the geometric position of the center of the electronic orbit up to an uncertainty arising from their noncommutativity. Therefore, the operators $\hat{b}$ and $\hat{b}^{\dagger}$ encode information about the center of the orbit. Since $\hat{X}_{\hat{b}}$ and $\hat{Y}_{\hat{b}}$ commute with $\hat{h}$, it follows that the Landau problem is invariant under spatial translations of the orbit center; that is, the energy does not depend on the position of this center \cite{Li1999PhysicaB, Kitadono2020GuidingCenter}. 

Using Eq.~\eqref{eq:Pi-b}, we express the guiding-center operators in terms of $\hat{\Pi}$ and $\hat{\Pi}^{\dagger}$ as
\begin{equation}
    \hat{X}_{\hat{\Pi}} = \frac{l_{B}^{2}}{2\hbar}\left(\hat{\Pi}^{\dagger}+\hat{\Pi}\right) \quad \text{and} \quad \hat{Y}_{\hat{\Pi}} = \ii \frac{l_{B}^{2}}{2\hbar}\left(\hat{\Pi}^{\dagger}-\hat{\Pi}\right).
\end{equation}
The coherent character of the states $\psi_n$ ensures that the guiding center is localized with the minimum uncertainty allowed by its noncommuting coordinates,
\begin{equation}
    \Delta X_{\hat{\Pi}}\,\Delta Y_{\hat{\Pi}}
    = \frac{l_B^2}{2}.
    \label{eq:guiding-center-uncertainty}
\end{equation}
These operators do not commute with the Hamiltonian $\hat{H}$ of the system,
\begin{equation}
    \left[\hat{X}_{\hat{\Pi}},\hat{H}\right] =  l_{B}^{2}e\mathcal{E} = 
    -\ii\left[\hat{Y}_{\hat{\Pi}},\hat{H}\right].
\end{equation}
Although the two self-adjoint guiding-center components do not separately commute with the Hamiltonian, their complex combination,
\begin{equation}
    \hat{\Pi}=\frac{\hbar}{l_B^2}
    \left(\hat{X}_{\hat{\Pi}}+\ii\hat{Y}_{\hat{\Pi}}\right),
\end{equation}
does commute with $\hat{H}$, whereas its adjoint does not. The perturbation therefore preserves a non-self-adjoint guiding-center generator.

\begin{figure*}[t]
  \centering
  \includegraphics[scale=0.30]{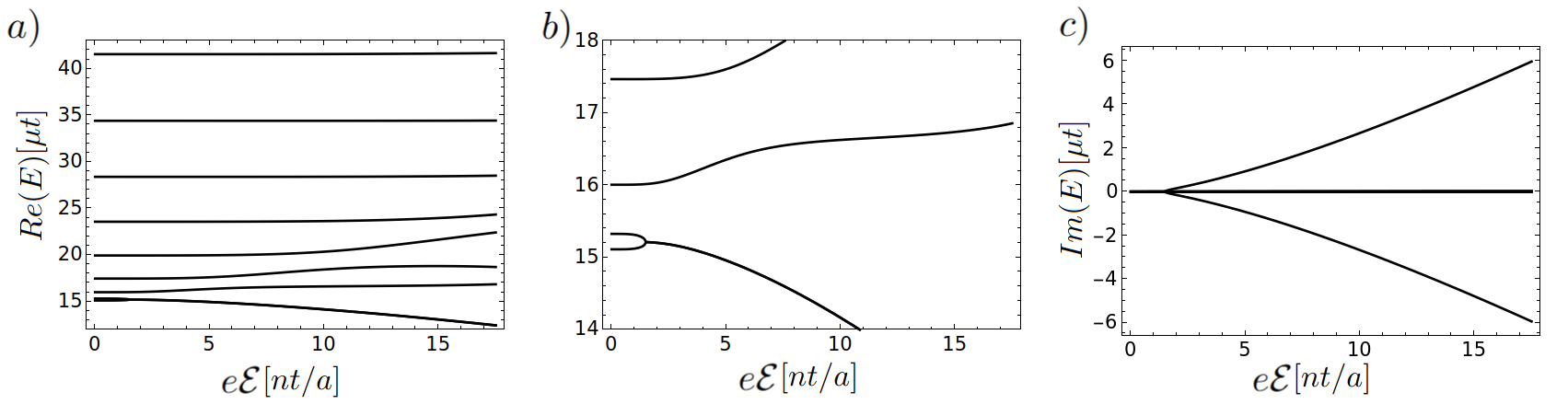}
  \caption{Low-energy spectrum of a tight-binding square lattice of $2000 \times 2000$ sites at $\phi=2.4\,[\mu a^{-2}]$ as a function of $e{\cal E}$. $(a)$ and $(b)$ panels show the real parts of the first nine eigenvalues and a zoom respectively. ($c$) Imaginary parts of them. Nonzero imaginary parts emerge in pairs with opposite signs after the corresponding eigenvalues coalesce at the exceptional point otherwise all the eigenvalues have null imaginary part.}
  \label{fig:eigenvalues}
\end{figure*}

The expectation values of the guiding-center operators are computed in Appendix \ref{app:XY-expected}, from which the physical meaning of $\zeta$ becomes transparent, leading to the following expression 
\begin{equation}
  \braket{\hat{X}_{\hat{\Pi}} } = 2l_{B}^{2} \zeta_{R}, \quad 
  \text{and} \quad
  \braket{\hat{Y}_{\hat{\Pi}} } = 2l_{B}^{2}\zeta_{I},
\end{equation}  
where $\zeta=\zeta_{R}+\ii\zeta_{I}$. This means that the position of the orbit center is 
\begin{equation}
  P_{c} = \left(\braket{\hat{X}_{\hat{\Pi}} }, 
  -\braket{\hat{Y}_{\hat{\Pi}} } \right)
  = 2l_{B}^{2}\left(\zeta_{R}, -\zeta_{I} \right).
\end{equation}

The centroid of the probability density is displaced relative to the position of the orbit center by an amount that depends on $n$,
\begin{equation}
\begin{aligned}
    &\left(\braket{\hat{x}},\,\braket{\hat{y}}\right)
    = P_c+(\delta_n,0),\\
    \delta_n
    =\kappa & \left[
    1+\frac{L_{n-1}^{(1)}(-\gamma^2/4)}
    {L_n(-\gamma^2/4)}
    \right],\qquad n\geq1,
\end{aligned}
\label{eq:density-centroid}
\end{equation}
with $\delta_0=\kappa$ and $\kappa=me\mathcal{E}/[2\hbar^2(2\alpha)^2]$.

To relate the gain and loss induced by the perturbation to the imaginary part of the energy, we decompose the Hamiltonian as $\hat{H}=\hat{H}_{H}+\hat{U}$, where $\hat{H}_{H}$ is the Hermitian part and $\hat{U}=\ii e\mathcal{E}\hat{y}$ is the non-Hermitian term.
The adjoint operator can be written as $\hat{H}^{\dagger} = \hat{H} - 2 \hat{U}$, leading to the relation:
\begin{equation}
\braket{\hat{H}^{\dagger}\psi_{n}| \psi_{n}} = E_{n}^{\ast} \braket{\psi_{n} | \psi_{n}} + 2\ii e \mathcal{E} \braket{\psi_{n} | \hat{y} | \psi_{n}},
\end{equation}
while for $\hat{H}$ we have:
\begin{equation}
\braket{\psi_{n} | \hat{H} \psi_{n}} = E_{n} \braket{\psi_{n}|\psi_{n}}.
\end{equation}
Equating these expressions by the adjoint identity gives, for normalized eigenstates,
\begin{equation}
    e\mathcal{E} \braket{\psi_{n} | \hat{y} | \psi_{n}} = \text{Im}(E).
    \label{eq:condition}
\end{equation}
By calculating the expectation value $\braket{\psi_{n} | y|\psi_{n}}$ (see Appendix \ref{app:Expected-value}), we find that it is equal to 
\begin{equation}
\braket{\psi_{n}| \hat{y} | \psi_{n}} = - \frac{\zeta_{I}}{2\alpha}, \quad \text{with} \quad \zeta_{I} = \text{Im}(\zeta). 
\label{eq:balance}
\end{equation}
Together with Eq.~\eqref{eq:energies}, this result verifies Eq.~\eqref{eq:condition} for every $\zeta$.
Furthermore, Eq.~\eqref{eq:balance} shows that the eigenvalues associated with states satisfying $\braket{\hat{y}}=0$ are real. 
In this configuration, the imaginary potential induces balanced net gain and loss. In particular, restricting $\zeta$ to real values selects this sector and positions the orbit center along the $x$-axis.

To illustrate the relation between real eigenvalues and balanced gain and loss in the finite system, we calculate the low-energy spectrum of a tight-binding square lattice of $2000 \times 2000$ sites with $\phi=2.4\,[\mu a^{-2}]$. Figure~\ref{fig:eigenvalues}$(a)$ and $(b)$ shows the real parts of the first few eigenvalues, whereas Fig.~\ref{fig:eigenvalues} ($c$) displays the imaginary parts of them. As $e{\cal E}$ increases, pairs of initially real eigenvalues coalesce at the exceptional points and then develop imaginary parts of opposite sign. The relation $\operatorname{Im}(E_n)=e\mathcal{E}\langle\hat{y}\rangle_n$ Eq.~\eqref{eq:condition}, which also holds for the lattice eigenstates, implies that each real eigenvalue corresponds to a state with balanced net gain and loss. In contrast to Fig. (\ref{fig:nanoribbon_dos}), Fig. (\ref{fig:eigenvalues}) does not show Landau levels because the spectrum is calculated in real space, where bulk states are mixed with edge states, therefore these levels are not expected to emerge.

\begin{figure*}[t]
  \centering
  \includegraphics[scale=0.29]{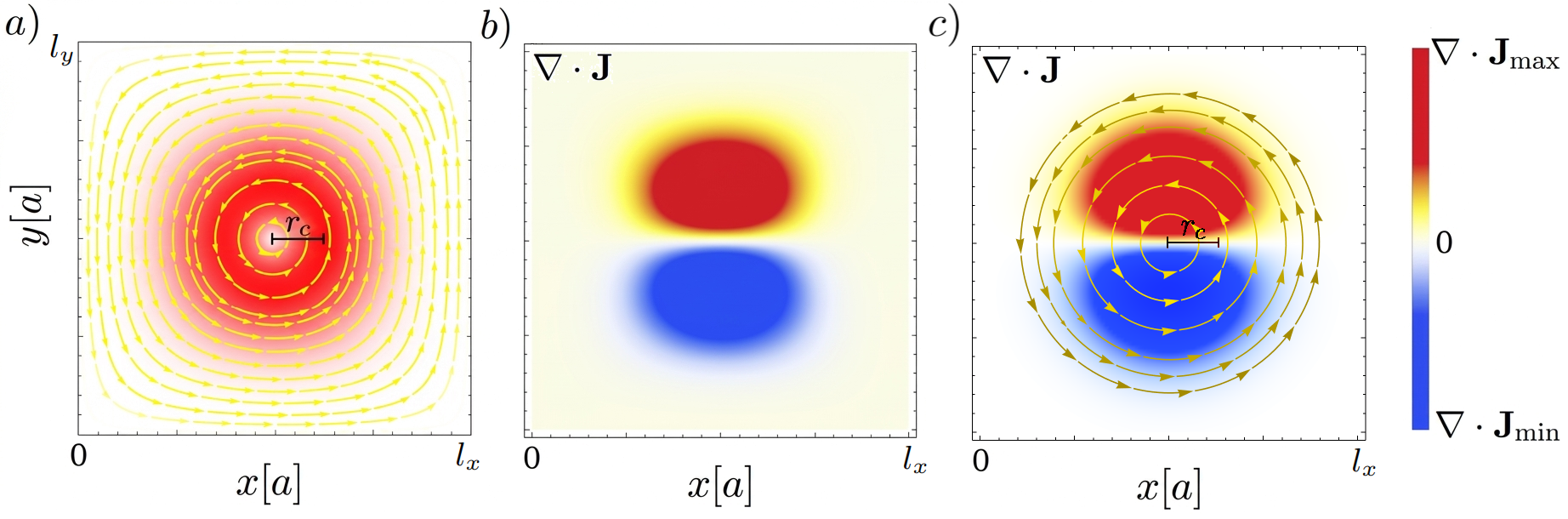}
  \caption{Probability current, Eq.(\ref{usual_current}), and its divergence for the lowest-energy eigenstate of the non-Hermitian Hamiltonian with $\phi=2.4[\mu a^{-2}]$ and $e{\cal E}=1 \,[nt/a]$, on a square sample of $l_{x}=l_{y}=2000 [a]$ sites. Panel $a)$, orbit described by the particle in the lowest energetic level, the horizontal line at the center of the lattice indicates the semiclassical estimated radius $r_{c}=257.5[a]$. Panel $b)$, divergence, Eq.(\ref{diver_0}), of the current shown in the previous panel. The gain region acts as a source (red lobe) and the loss region as a sink (blue lobe), Due to the symmetry of these regions, the gains and losses remain balanced leading to a real spectrum. Panel $c)$, circular patterns described by the particle calculated with the analytical expression Eq.(\ref{electric_current}) for $n=0$, and the divergence Eq.(\ref{diver_0}) calculated with the analytic wave function.}
  \label{fig:orbita_divergencia}
\end{figure*}
\section{\label{sec:current_for_all_n}Electric current and resistance}
The current of the system for the lowest Landau level, with energy $E_0$, can be visualized using the eigenvectors of the Hamiltonian Eq.(\ref{TBH2}), we denote these vectors by 
\begin{equation}
    \ket{\psi}=\sum_{i}\psi_{i}\ket{i},
\end{equation}
such that they satisfy the eigenvalue equations
\begin{equation}
    {\hat H}\ket{\psi}=E_{0}\ket{\psi}.
\end{equation}
For the tight-binding model, the local current from $i$ to $j$ atoms can be calculated using the expression
\begin{equation}\label{usual_current}
    J=\frac{e \ii}{\hbar}\left(t_{ij}\psi_{i}^{*}\psi_{j}-t_{ji}\psi_{j}^{*}\psi_{i}\right).
\end{equation}
where $t_{ij}$ are the elements of the Hamiltonian matrix \ref{TBH2}, $\psi_{i}$ are the component of the eigenvectors. This current obeys a balance equation with gain and loss. The probability density in the $i$-th site as $n=|\psi_i|^2$, is given by the following expression
\begin{equation}\label{continuity2}
    e\frac{\partial n}{\partial t}
    +\nabla\cdot\mathbf{J}
    =\frac{2e}{\hbar}e\mathcal{E}y|\psi|^2.
\end{equation}
Here, the lattice divergence is evaluated as the sum of the outgoing bond currents at each site. For the eigenstate considered, Eq.~\eqref{continuity2} gives the divergence,
\begin{equation}\label{diver_0}
    \nabla\cdot\mathbf{J}
    =\frac{2e}{\hbar}
    \left(e\mathcal{E}y-\mathrm{Im}(E_0)\right)|\psi|^2.
\end{equation}
For real eigenvalues, positive and negative divergence correspond directly to local gain and loss, respectively. For complex eigenvalues, the term involving $\mathrm{Im}(E_0)$ accounts for the overall temporal growth or decay of the probability density.

Figure~\ref{fig:orbita_divergencia} displays the probability current patterns for $e{\cal E} = 1\,[nt/a]$ and $\phi = 2.4[\mu a^{-2}]\,$ in a square-lattice sample of $N = 2000 \times 2000$ sites. Panel $a)$ shows the trajectory followed by the particle, the red central region indicates the highest probability density, while the streamlines depict the circulating current pattern. Panel $b)$ illustrates the divergence of the current calculated via Eq.~\eqref{diver_0}, where positive values (red) correspond to a source (gain region) and negative values (blue) correspond to a sink (loss region). Panel $c)$, depicts the current pattern using the analytic solution Eq.(\ref{waave_func}) for $n=0$ calculated with the equation Eq.(\ref{electric_current}) and its respective divergence given by Eq.(\ref{diver_0}) using the respective wave function. A good agreement with the tight-binding calculations can be observed.
A semiclassical estimation of the cyclotron radius can be obtained as follows. The radius in terms of the energy $E$ is given by $r_{c}=\sqrt{2mE}/eB$. By substituting the effective mass relation derived in Section~\ref{sec:solution} and setting the energy equal to that of the unperturbed lowest Landau level ($E=\hbar\omega_{c}/2$), the radius can be expressed as $r_{c}=1/\sqrt{2\pi\phi}$. Forthis case the estimated value of the radius is $r_{c}=257.5~[a]$.

For the continuum model, we calculate the electric current using
\begin{equation}\label{electric_current}
{\bf J}_{n}=\frac{i e\hbar}{2m}\left(\psi_{n}\nabla\psi_{n}^{*}-\psi_{n}^{*}\nabla\psi_{n}\right)-\frac{e^{2}}{m}\vec{A}|\psi_{n}|^{2}.
\end{equation}
Substituting the Eq.(\ref{waave_func}) and making the following change of variable
\begin{align}
  \xi_{r}\cos\xi_{\theta} &= \sqrt{2\alpha}\left(x - \frac{\zeta_{R}}{2\alpha}\right),  \\
  \xi_{r}\sin\xi_{\theta} &= \sqrt{2\alpha}\left(y + \frac{\zeta_{I}}{2\alpha}\right),
\end{align}
one can obtain the expression
\begin{equation}\label{current_n}
    {\bf J}_{n}=e{\bf v_{n}}|\psi_{n}|^{2},
\end{equation}
where the velocity vector ${\bf v}_{n}=(v_{n_{x}},v_{n_{y}})$ has coordinates
\begin{align}
  v_{n_{x}} &= \frac{\omega_{c}}{\sqrt{8\alpha}}\left(\xi_{r}+\frac{n}{\xi_{r}}\right)\sin\xi_{\theta}, \label{eq:current-compX} \\
    v_{n_{y}} &= -\frac{\omega_{c}}{\sqrt{8\alpha}}\left[\left(\xi_{r}+\frac{n}{\xi_{r}}\right)\cos\xi_{\theta}+\frac{\gamma}{2}\right], \label{eq:current-compY}
\end{align}
where $\gamma$ is defined by Eq.(\ref{eq:gamma}). 
In these shifted coordinates, both the normalized probability density and the velocity field are independent of $\zeta$. Consequently, their probability-weighted spatial average over the entire plane has the same form for every complex $\zeta$.
We then evaluate the probability-weighted mean current density,
\begin{equation}
  \braket{{\bf J}_{n}}=e(\bra{\psi_{n}^{2}}v_{n_{x}}\ket{\psi_{n}^{2}},\bra{\psi_{n}^{2}}v_{n_{y}}\ket{\psi_{n}^{2}}). \label{eq:current-field}
\end{equation}
See Appendix \ref{app:current-vector} for details of the calculation.
By symmetry of Eq.(\ref{eq:current-compX}) with respect to $\xi_{\theta}$, we find that 
\begin{equation}
  \bra{\psi_{n}^{2}}v_{n_{x}}\ket{\psi_{n}^{2}} = 0.
\end{equation}
The $Y$-component of the probability-weighted mean current density is determined by 
\begin{align}
  \bra{\psi_{n}^{2}}v_{n_{y}}\ket{\psi_{n}^{2}} &= -
  \frac{(\frac{1}{2})_{n}}{n!}\frac{e\mathcal{E}}{2\pi\hbar}
  \frac{L_{2n}^{(1)}\left(-\frac{\gamma^{2}}{2}\right)}{\left[L_{n}\left(-\frac{\gamma^{2}}{4}\right)\right]^{2}},\label{eq:drift_vy}
\end{align}
\begin{figure}[t]
  \centering
  \includegraphics[scale=0.38]{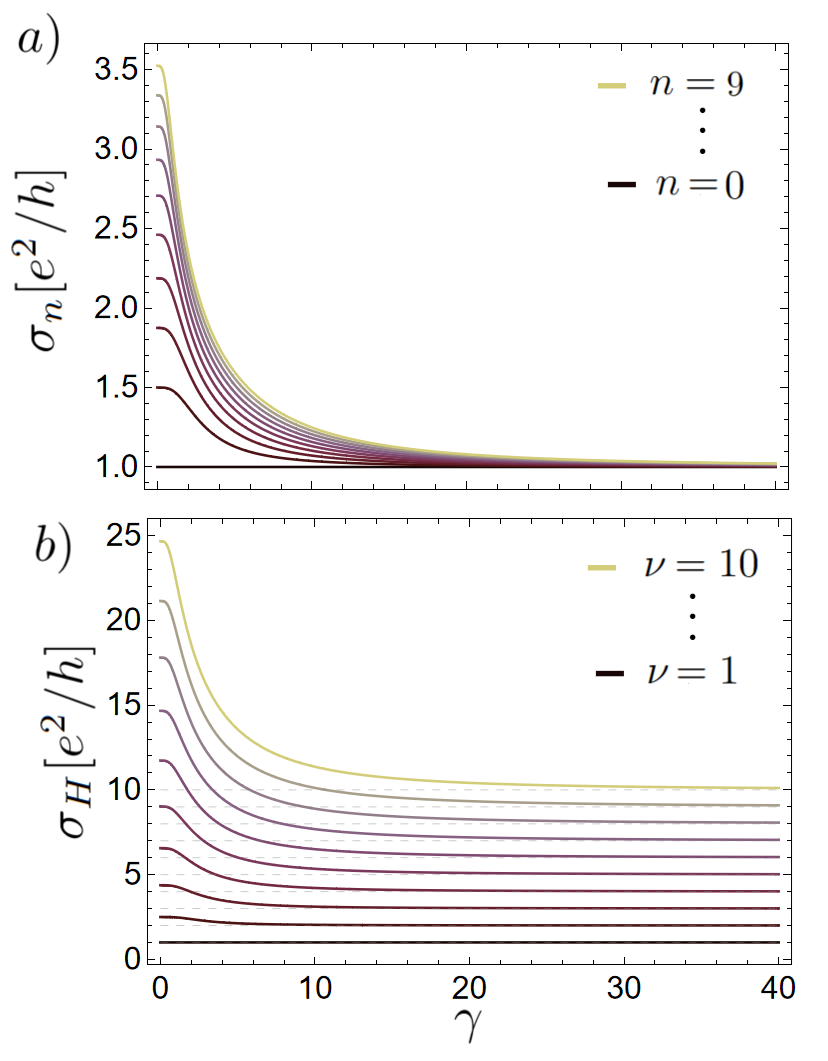}
  \caption{Conductivity per Landau level obtained from Eqs.~\eqref{eq:currentDensity} and \eqref{eq:levelsFill}. (a) Contributions $\sigma_n$, in units of $e^2/h$, for $n=0,\ldots,9$. The $n=0$ contribution is exactly unity, whereas each fixed $n>0$ approaches unity as $\gamma\to\infty$. (b) Total Hall conductivity within the independent-level approximation for $\nu=1,\ldots,10$ completely filled levels. The curves approach the integer values $\nu$, indicated by dashed lines.}
  \label{fig:Hall_quantization}
\end{figure}
where $(a)_{n}$ denotes the Pochhammer symbol and $L_{n}^{(1)}(x)$ is the first associated Laguerre polynomial. From Eq.~\eqref{eq:drift_vy}, the magnitude of the probability-weighted mean transverse current density per unit electric field defines the response coefficient associated with the normalized state $\psi_n$,
\begin{equation}
\sigma_n=
\frac{e^2}{h}\frac{(\frac12)_n}{n!}
\frac{L_{2n}^{(1)}\left(-\frac{\gamma^2}{2}\right)}
{\left[L_n\left(-\frac{\gamma^2}{4}\right)\right]^2}.
\label{eq:currentDensity}
\end{equation}

To relate this single-state calculation to the response of the electron ensemble, we establish an effective transport framework. We assume that $\sigma_n$ represents the effective transverse contribution of the corresponding filled level, with the occupation of its guiding-center sector already accounted for. Neglecting electron--electron interactions, spin degeneracy, and scattering between different levels, we consider the levels $n=0,\ldots,\nu-1$ to be filled. Their currents add independently under the same electric field, giving
\begin{equation}
\sigma_H=\sum_{n=0}^{\nu-1}\sigma_n,
\label{eq:levelsFill}
\end{equation}
where $\nu$ is the integer filling factor and counts the filled levels.

Figure~\ref{fig:Hall_quantization}(a) shows the contributions $\sigma_n$ for $n=0,\ldots,9$, whereas panel (b) displays their sums for $\nu=1,\ldots,10$. The contribution $\sigma_0$ equals $e^2/h$ for all $\gamma$. For each fixed $n>0$, $\sigma_n$ approaches the same value as $\gamma\to\infty$. Consequently, within the effective description adopted above, $\sigma_H$ approaches the integer multiples $\nu e^2/h$.

The time-dependent wave function associated with each eigenstate is
\begin{equation}\label{eq:time_wf0}
    \psi_{n}(\mathbf{x},\tau)
    = \psi_{n}(\mathbf{x})
    \exp\!\left[-\ii\frac{\mathrm{Re}(E_n)}{\hbar}\tau\right]
    \exp\!\left[\frac{\mathrm{Im}(E_n)}{\hbar}\tau\right],
\end{equation}
where $\tau$ is the time variable. At each time, the spatial average is taken with the normalized probability density, while the local current density is evaluated from the unnormalized evolving state. Under this convention, the oscillatory phase cancels, yielding
\begin{equation}
    \sigma_n(\tau)
    = \sigma_n
    \exp\!\left[\frac{2\mathrm{Im}(E_n)}{\hbar}\tau\right],
\end{equation}
where $\sigma_n$ is the initial response coefficient given by Eq.~\eqref{eq:currentDensity}.

\begin{figure*}[t]
  \centering
  \includegraphics[scale=0.35]{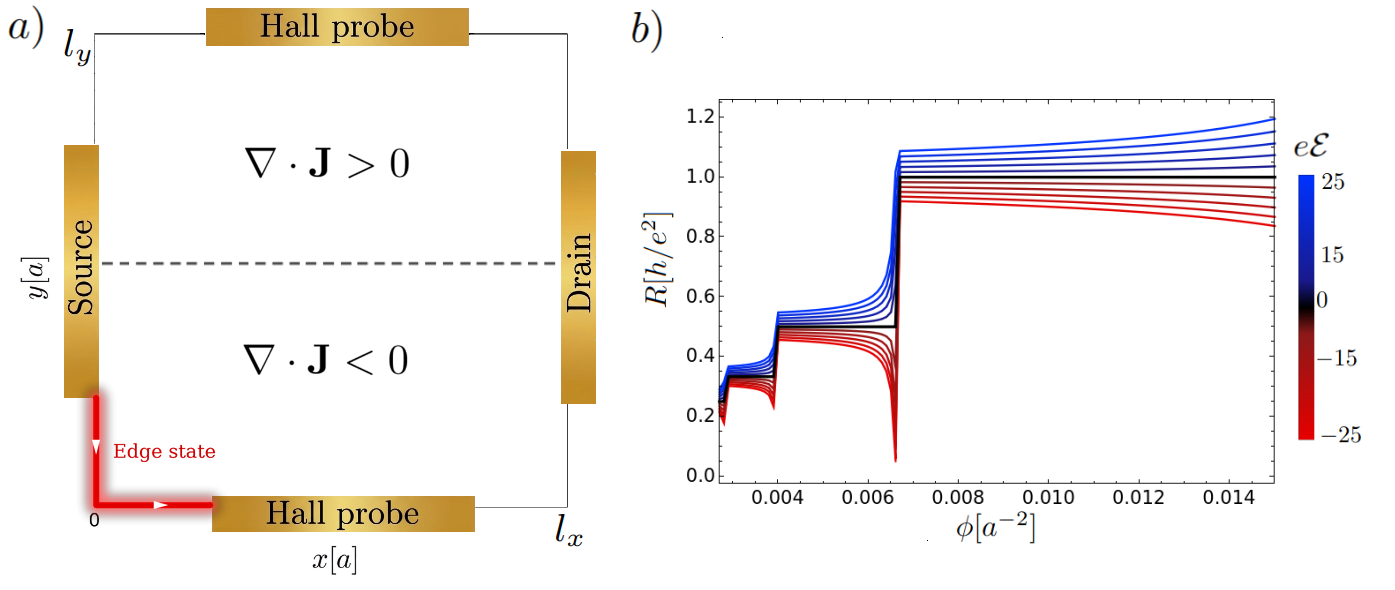}
  \caption{Panel $a)$, schematic setup of a square-lattice with source and drain connected to the left and right sides of the system and Hall probes connected to the top and bottom, respectively. An edge state observed at the bottom left corner for a magnetic flux density $\phi=0.008~[a^{-2}]$ for a $2000\times2000$ square lattice for an electron injected at energy $E = -3.876\,[t]$ and $e{\cal E}=15~[nt/a]$. The dashed line indicates the middle of the sample that divides the positive divergence region from the negative one. Panel $b)$, Hall resistance $R$, in units of $h/e^2$, as a function of $\phi$. The perturbation strength $e{\cal E}$ is given in units of $\text{n} t / a$. Positive values of $e{\cal E}$ (blue curves), show an upward linear shift of the Hall plateaus. Negative values of $e{\cal E}$ (red curves), where the plateaus shift downward. In both cases, the plateau shifts scale linearly with a rate $\mathcal{C} = 0.02/e{\cal E}$. When $e{\cal E}=0$ we obtain the usual Hall quantization (black curve).}
  \label{fig:Hall_resistance}
\end{figure*}

When the quantized regime is approached,
\begin{equation}\label{eq:Hall-quantization}
    \lim_{\gamma\to\infty}\sigma_n=\frac{e^2}{h},
\end{equation}
the time-dependent conductivity takes the asymptotic form
\begin{equation}\label{eq:Hall-quantization-time}
    \sigma_n(\tau)\simeq
    \frac{e^2}{h}
    \exp\left(\frac{2\mathrm{Im}(E_n)}{\hbar}\tau\right).
\end{equation}
For $n=0$, this expression is exact at any $\gamma$. According to Eq.~\eqref{eq:energies}, $\mathrm{Im}(E_n)=-e\mathcal{E}\zeta_I/(2\alpha)$. Therefore, for $e\mathcal{E}>0$ and $\alpha>0$, decay requires $\zeta_I>0$, whereas growth requires $\zeta_I<0$. At fixed $\zeta$ and system parameters, the exponential rate is the same for all $n$. Finally, using $L_n(0)=1$ and $L_{2n}^{(1)}(0)=2n+1$, the opposite limit of the initial response coefficient is
\begin{equation}
    \lim_{\gamma\to0}\sigma_n
    =\frac{e^2}{h}\left(\frac{1}{2}\right)_n
    \frac{(2n+1)}{n!}.
\end{equation}


To numerically study the resistance of this system, we employed Non-Equilibrium Green's Functions (NEGF); for the computational details, see Appendix \ref{app:NEGF}.
Our setup is depicted in Fig.~\ref{fig:Hall_resistance} ($a$). Four contacts are connected to the sample: a source and a drain at the left and right boundaries, respectively, along with two Hall probes connected at the top and bottom edges. Besides, it shows the edge state formed at the bottom left corner when an electron is injected into the system with an energy of $E = -3.876 \,[t]$, $\phi=0.008[a^{-2}]$, close to the lowest Landau level, see Fig. \ref{fig:nanoribbon_dos} and $e{\cal E}=15~[nt/a]$ in a square lattice with length of $2000$ atoms. As can be seen, the current goes directly from the source to the inferior Hall probe going through the region where the divergence is negative. In Fig.~\ref{fig:Hall_resistance} ($b$), we show the calculated resistance between the Hall probes as a function of $\phi$. It can be observed that the curves shift as the parameter $e{\cal E}$ changes, leading to linear displaced Hall plateaus.

The observed shift of the resistance can be interpreted in terms of attenuation and intensification induced by the non-Hermitian perturbation. When the state is an edge state going through the region where $\nabla\cdot{\bf J}<0$, the $\mathrm{Im}(E)<0$ and the exponential factor in Eq.~\eqref{eq:Hall-quantization-time} describes a decrease of the current during propagation. From Eq.~\eqref{eq:energies}, its exponent is proportional to $-\beta$. Over the plateau interval, we assume that the combination $\zeta_I\tau/\omega_c$ can be treated as approximately constant (considering that the time it takes the electron to go from the source to the Hall probe is finite), allowing this factor to be expressed as $\exp(-\mathcal{C}\beta)$, with an effective positive coefficient $\mathcal{C}$.

For the Hall plateau quantized exactly when $R=h/e^{2}$, expanding this factor to first order in $\mathcal{C}\beta\ll1$ gives $\sigma_H\approx(e^2/h)(1-\mathcal{C}\beta)$. The effective relation $R\approx1/\sigma_H$ for the measurement configuration considered leads to a Hall resistance of the form 
\begin{equation}
    R \sim \frac{h}{e^{2}}\frac{1}{1-\mathcal{C}\beta}
    \sim \frac{h}{e^{2}}(1+\mathcal{C}\beta),
\end{equation}
leading to the shift observed in the blue curves in Fig. \ref{fig:Hall_resistance} ($b$). Red curves can be understood similarly than the previous case, but changing the sign of $e{\cal E}$, which leads to the same above expression but with a negative slope respect this parameter. Hence the linear displacement can be written $R\sim h(1-\mathcal{C}\beta)/e^{2}$, naturally this causes the current to increase because the region with negative divergence shift with the positive one leading to a diminution of the Hall plateaus. The value of the displacement for both cases is equal to $\mathcal{C}=0.02/e{\cal E}$, being $e{\cal E}\neq 0$. For the case when $e{\cal E}=0$ the perturbation is not being considered and the usual Hall quantization is recovered, this situation is illustrated by the black curve.

\section{\label{sec:conclusions}Conclusions}

We investigated electronic states and Hall transport in a square lattice subject to a uniform magnetic field and a non-Hermitian linear perturbation. The analytical continuum solution yields energy branches with Landau spacing at fixed $\zeta$ and a complex linear Stark shift. The separation constant $\zeta$ labels coherent eigenstates of the conserved non-Hermitian operator $\hat{\Pi}$ and determines the orbit center. Although the two self-adjoint guiding-center components do not separately commute with the Hamiltonian, their complex combination $\hat{\Pi}$ does. For real $\zeta$, the net gain and loss are balanced and the corresponding energies are real for any perturbation strength within the continuum model.

The finite-lattice calculations support the Landau-level structure and the relation between real eigenvalues and balanced net gain and loss. Specifically, they show a particle orbit centered along the $x$-axis of the lattice, with its upper section in the gain region and its lower section in the loss region. In the low magnetic field intensity regime, as the perturbation increases, pairs of initially real eigenvalues coalesce and acquire imaginary parts of opposite signs, consistent with the formation of exceptional points. These coalescences are characteristic of the finite-sample spectrum.

Within the effective transport description adopted here, the contribution associated with $n=0$ is exactly $e^2/h$, while each fixed $n>0$ approaches this value as $\gamma\to\infty$. For a fixed number $\nu$ of filled levels, the assumed additive response therefore approaches $\nu e^2/h$. The numerical transport calculation, performed at the fixed injection energy $E=-3.876\,[t]$, which is close to the lowest Landau level, shows a shift of the Hall resistance plateaus above or below $h/e^2$ depending on the sign of $e{\cal E}$.

These results motivate a systematic study of how sample size, boundary geometry, and contact coupling affect the displacement of the Hall plateau and the quantitative agreement between the continuum and lattice descriptions. A microscopic treatment of the gain and loss processes would further help connect the effective perturbation to an experimental implementation. Recent circuit demonstrations of non-Hermitian Landau modes~\cite{wang2026reshaping,zhang2024continuum} also motivate the exploration of analogous wave systems, where complex linear potentials could be used to control mode localization and spatial profiles.

\section*{Acknowledgment}
This work was supported by the DGAPA-UNAM Postdoctoral Program (POSDOC), UNAM-PAPIIT project No. IG100725 and project SECIHTI Fronteras CF-2023-G-763. K.U. acknowledges financial support from SECIHTI through the program ``Estancias Posdoctorales por M\'exico''. S.G. is a doctoral student from the Posgrado en Ciencias (Física) at UNAM and acknowledges the financial support provided by the Secretaría de Ciencia, Humanidades, Tecnología e Innovación (SECIHTI) through the doctoral fellowship (CVU No. 1103995).

\appendix
\section{\label{app:normalization} Wave function normalization}

The inner product of the wave function can be written as  
\begin{equation}
  \braket{\psi_{n} | \psi_{n}} = |A_{n}|^{2}(2\alpha)^{n-1}\exp\! \left(\frac{|\zeta|^{2}}{2\alpha}\right)
  \Upsilon_{0}. \label{eq:app_norma-upsilons}
\end{equation}
where
\begin{align}
  \Upsilon_{0} &= \int\limits_{0}^{\infty}\int\limits_{0}^{2\pi}\exp(-\xi_{r}^{2}
  +\gamma\xi_{r}\cos\xi_{\theta} )\xi_{r}^{2n+1} d\xi_{\theta}d\xi_{r} \label{eq:app_upsilon0} 
\end{align}
where the changes of variables,
\begin{align}
  \xi_{r}\cos\xi_{\theta} &= \sqrt{2\alpha}\left(x-\frac{\zeta_{R}}{2\alpha}\right), \label{eq:app_xix}\\ 
  \xi_{r}\sin\xi_{\theta} &= \sqrt{2\alpha}\left(y+\frac{\zeta_{I}}{2\alpha}\right), \label{eq:app_xiy}\\
  \xi_{r}d\xi_{r}d\xi_{\theta} &= 2\alpha dxdy, \label{eq:app_difs-xis}
\end{align}
were made, and we defined
\begin{equation}
  \kappa=\frac{me\mathcal{E}}{2\hbar^{2}(2\alpha)^{2}} \quad \text{and} \quad 
  \zeta = \zeta_{R} + \ii \zeta_{I}. \label{eq:app_def-k-z}
\end{equation}
such that $\gamma = 2\kappa\sqrt{2\alpha}$. 
To evaluate these integrals, we make use of the relation between the modified Bessel functions of the first kind and the ordinary Bessel functions, $I_{\nu}(x)=\ii^{-\nu}J_{\nu}(\ii x)$; as well as the formulas (\cite{gradshteyn2007}, Eq.8.431.3) and (\cite{gradshteyn2007}, Eq.6.631.1), 
\begin{equation}
  I_{\nu}(z) = \frac{(z/2)^{\nu}}{\Gamma(\nu+1/2)\Gamma(1/2)} \int_{0}^{\pi}e^{\pm z \cos\theta}\sin^{2\nu}\theta d\theta \label{eq:app_clave1}
\end{equation}
\begin{align}
  \int_{0}^{\infty}z^{\mu}e^{-z^{2}}I_{\nu}(\lambda z)dz &=
  \frac{\lambda^{\nu}\Gamma\left[\frac{1}{2}(\nu+\mu+1)\right]}
  {2^{\nu+1}\Gamma(\nu+1)} \nonumber \\
  & \quad \times {}_{1}F_{1}\left(\frac{\nu+\mu+1}{2},\nu+1,\frac{\lambda^{2}}{4}\right) \label{eq:app_clave2}
\end{align}
Thus, for $\Upsilon_{0}$ one obtains
\begin{align}
  \Upsilon_{0} &= 2\pi \int\limits_{0}^{\infty}\exp(-\xi_{r}^{2})\xi_{r}^{2n+1} I_{0}(\gamma\xi_{r})d\xi_{r} \nonumber \\ 
  &= \pi\Gamma(n+1)\,{}_{1}F_{1}\left(n+1,1,\frac{\gamma^{2}}{4}\right) \label{eq:app_upsilon0-resuelto}.
\end{align}
Hence, by imposing the normalization condition $\braket{\psi_{n} | \psi_{n}} = 1$, it follows that 
\begin{equation}
  |A_{n}|^{2}\ = \frac{(2\alpha)^{1-n}\exp\! \left(-\frac{|\zeta|^{2}}{2\alpha}\right)}{\pi\Gamma(n+1)\,{}_{1}F_{1}\left(n+1,1,\frac{\gamma^{2}}{4}\right)}.
\end{equation}
This expression can be written in terms of the Laguerre polynomial $L_{n}(x)$ using of Kummer's identity (\cite{gradshteyn2007}, Eq.9.212.1), 
\begin{equation}
  {}_{1}F_{1}\left(a,b,x \right) = e^{x}{}_{1}F_{1}\left(b-a,b,-x\right), \label{eq:app_clave3}
\end{equation}
as well as the relation between confluent hypergeometric functions and Laguerre polynomials (\cite{gradshteyn2007}, Eq.8.972.1),
\begin{equation}
  L_{\nu}^{\eta}(x) = \binom{\nu+\eta}{\nu}{}_{1}F_{1}\left( -\nu, \eta+1,x\right). \label{eq:app_clave4}
\end{equation}
These identities lead directly to the expression
\begin{equation}
  |A_{n}|^{2}\ = \frac{(2\alpha)^{1-n}\exp\! \left(-\frac{|\zeta|^{2}}{2\alpha}\right)}{\pi n!\exp\left(\frac{\gamma^{2}}{4}\right)L_{n}\left(-\frac{\gamma^{2}}{4}\right)}.
\end{equation}

\section{\label{app:Symmetries} Symmetry operator}

In this Appendix, we derive the linear differential operator that represents the first-order symmetries of the Hamiltonian \cite{Miller1977Symmetry}.
By expressing the Hamiltonian of the equation (\ref{Heff_z}) in differential form, it takes the form
\begin{equation}
  \hat{H} = a(\partial_{xx} + \partial_{yy}) +b(x\partial_{y}-y\partial_{x})+c(x^{2}+y^{2})+d(x-\ii y), \label{eq:app_hamiltoniano-diff}
\end{equation}
with
\begin{equation}
    a=-\frac{\hbar^{2}}{2m}, \quad b=\frac{\ii\omega_{c}\hbar}{2}, \quad 
    c=\frac{m\omega_{c}^{2}}{8}, \quad d=-e\mathcal{E},
\end{equation}
and the symmetry operator is expressed as 
\begin{equation}
  \hat{\Pi}=X\partial_{x}+Y\partial_{y}+Z, \label{eq:app_pi-diff}
\end{equation}
with $X=X(x,y)$, $Y=Y(x,y)$ and $Z=Z(x,y)$.
Substituting Eqs. (\ref{eq:app_hamiltoniano-diff}) and (\ref{eq:app_pi-diff}) into the commutation relation $\left[\hat{\Pi},\hat{H}\right]=0$, and collecting terms of the same order, one finds the following system of equations
\begin{align}
        &  X_{x}=Y_{y}=0, \quad X_{y}+Y_{x}=0, \label{eq:a}\\[5pt]
        & aX_{xx}+bY+aX_{yy}+2aZ_{x}+bxX_{y}-byX_{x} = 0, \nonumber \\
        & aY_{xx}-bX+aY_{yy}+2aZ_{y}+bxY_{y}-byY_{x} = 0, \label{eq:b}\\[5pt]
        & a(Z_{xx}+Z_{yy})+b(xZ_{y}-yZ_{x})\nonumber\\
        &-(2xc+d)X-(2yc-\ii d)Y=0. \label{eq:c}
\end{align}
From equation (\ref{eq:a}) one obtains $X=\eta y + e_1$ and $Y=-\eta x + e_{2}$, with $\eta$ constant. Substituting these expressions into equation (\ref{eq:b}), it follows that $Z_{x}=-\frac{b e_{2}}{2a}$ and $Z_{y}=\frac{b e_{1}}{2a}$. These relations are then substituted into equation (\ref{eq:c}), leading to the equation
\begin{equation}
    \eta(y+ix)+(e_{1}-\ii e_{2}) = 0,
\end{equation}
which imposes the conditions $\eta = 0$ and $e_{1} = \ii e_{2}$. Therefore,
\begin{equation}
    X=e_{1}, \quad Y=-\ii e_{1} \quad \text{and} \quad Z=2\alpha e_{1} (x-\ii y)+e_{3}.
\end{equation}
Without loss of generality, one may choose $e_{1} = \hbar$ to preserve the units of momentum, 
and the constant $e_{3}$ can be neglected since it represents the trivial action of the identity operator. In this way, we find that the symmetry operator $\hat{\Pi}$ takes the form
\begin{equation}
    \hat{\Pi} = \ii \hat{p}_{x} + \hat{p}_{y} + 2\hbar \alpha (\hat{x}-\ii\hat{y}).
\end{equation}

\section{\label{app:XY-expected} Expectation value of guiding-center operators}

We evaluate the expectation values of $\hat{X}_{\hat{\Pi}}$ and $\hat{Y}_{\hat{\Pi}}$ using the standard inner product for arbitrary $\zeta\in\mathbb{C}$.
In terms of the position and momentum operators, the guiding-center operators take the form
\begin{align}
  \hat{X}_{\hat{\Pi}} &= \frac{l_{B}^{2}}{\hbar}
  \left(\hat{p}_{y} +2\alpha \hbar \hat{x}\right) \\
  \hat{Y}_{\hat{\Pi}} &= \frac{l_{B}^{2}}{\hbar}
  \left(\hat{p}_{x} - 2\alpha \hbar \hat{y}\right).
\end{align}
We first compute the expectation value of the guiding-center operator along the horizontal direction,
\begin{align}
  \braket{\hat{X}_{\hat{\Pi}} } &= l_{B}^{2} 
  \iint\limits_{\mathbb{R}^{2}} |\psi_{n}|^{2}
  \bigg\lbrace 2\alpha (x + \ii y) + \zeta
  -2\alpha\kappa \nonumber \\
  & \quad -n \left[\left(x-\frac{\zeta_{R}}{2\alpha}\right) 
  - \ii \left(y + \frac{\zeta_{I}}{2\alpha}\right) \right]^{-1} \bigg\rbrace
  dxdy, \label{eq:app_valor-esperado-X}
\end{align}
Then, performing the change of variables Eqs.~\eqref{eq:app_xix}, \eqref{eq:app_xiy}, \eqref{eq:app_difs-xis} and the definition Eq.~\eqref{eq:app_def-k-z} the expectation value can be recast in a more compact form
\begin{align}
  \braket{\hat{X}_{\hat{\Pi}} } &= |A_{n}|^{2}l_{B}^{2}(2\alpha)^{n-1}\exp\left( \frac{|\zeta|^{2}}{2\alpha}\right) \nonumber \\
  & \quad \times\Big[ \sqrt{2\alpha}\left(\Upsilon_{2} - n\Upsilon_{1} \right) 
  + 2(\zeta_{R}-\alpha\kappa) \Upsilon_{0} \label{eq:app_val-esp-upsilons}
\Big]
\end{align}
where
\begin{align}
  \Upsilon_{1} &= \int\limits_{0}^{\infty}\int\limits_{0}^{2\pi}\exp(-\xi_{r}^{2}
  +\gamma\xi_{r}\cos\xi_{\theta} + \ii \xi_{\theta})\xi_{r}^{2n} d\xi_{\theta}d\xi_{r} \label{eq:app_upsilon2} \\
  \Upsilon_{2} &= \int\limits_{0}^{\infty}\int\limits_{0}^{2\pi}\exp(-\xi_{r}^{2}
  +\gamma\xi_{r}\cos\xi_{\theta} +\ii\xi_{\theta} )\xi_{r}^{2n+2} d\xi_{\theta}d\xi_{r}. \label{eq:app_upsilon3}
\end{align}
With the aid of the formula (\cite{gradshteyn2007},Eq.8.431.5),
\begin{equation}
  I_{\nu}(z) = \frac{1}{\pi} \int_{0}^{\pi} e^{z\cos\theta}\cos{\nu\theta}\d\theta, \label{eq:app_clave5}
\end{equation}
valid for $\nu\in\mathbb{Z}$, one obtains 
\begin{align}
  \Upsilon_{1} &= 2\pi \int\limits_{0}^{\infty}\exp(-\xi_{r}^{2})
  \xi_{r}^{2n} I_{1}(\gamma\xi_{r}) d\xi_{r} \label{eq:app_upsilon2-2} \\
  \Upsilon_{2} &= 2\pi \int\limits_{0}^{\infty}\exp(-\xi_{r}^{2})
  \xi_{r}^{2n+2} I_{1}(\gamma\xi_{r}) d\xi_{r}. \label{eq:app_upsilon3-2}
\end{align}
Then, using Eq. (\ref{eq:app_clave2}), the above integrals can be evaluated as
\begin{align}
  \Upsilon_{1} &= \frac{\pi\gamma}{2}\Gamma(n+1)\,
  {}_{1}F_{1}\left(n+1,2,\frac{\gamma^{2}}{4}\right) \label{eq:app_upsilon2-3} \\
  \Upsilon_{2} &= \frac{\pi\gamma}{2}\Gamma(n+2)\,
  {}_{1}F_{1}\left(n+2,2,\frac{\gamma^{2}}{4}\right) \label{eq:app_upsilon3-3}
\end{align}
We now note that the difference $\Upsilon_{2}-n\Upsilon_{1}$ can be reduced, with the aid of the recurrence relation (\cite{gradshteyn2007}, Eq.9.212.3),
\begin{align}
  & \nu\,{}_{1}F_{1}(\nu+1,\eta+1,x) =\label{eq:app_clave6} \\
  & \qquad \qquad (\nu-\eta)\,{}_{1}F_{1}(\nu,\eta+1,x) 
  + \eta\,{}_{1}F_{1}(\nu,\eta,x), \nonumber
\end{align}
to the simplified form
\begin{align}
  \Upsilon_{2}-n\Upsilon_{1} &= \frac{\pi \gamma}{2}\Gamma(n+1)\,
  {}_{1}F_{1}\left(n+1,1,\frac{\gamma^{2}}{4} \right).
\end{align}
Thus, the following terms in Eq. (\ref{eq:app_val-esp-upsilons}) cancel each other,
\begin{equation}
  \sqrt{2\alpha}\left(\Upsilon_{2} - n\Upsilon_{1} \right) 
  -2\alpha\kappa \Upsilon_{0} =0,
\end{equation}
leading to the final result, 
\begin{equation}
  \braket{X}_{\hat{\Pi}} = 2l_{B}^{2}\zeta_{R}.
\end{equation}

With the aid of the preceding equations, the expectation value of the guiding-center operator along the vertical direction follows almost directly,
\begin{align}
  \braket{\hat{Y}_{\hat{\Pi}} } &= \ii l_{B}^{2}
  \iint\limits_{\mathbb{R}^{2}} |\psi_{n}|^{2}
  \bigg\lbrace 2\alpha (x + \ii y) - \zeta
  -2\alpha\kappa \nonumber \\
  & \quad -n \left[\left(x-\frac{\zeta_{R}}{2\alpha}\right) 
  - \ii \left(y + \frac{\zeta_{I}}{2\alpha}\right) \right]^{-1} \bigg\rbrace
  dxdy. \label{eq:app_valor-esperado-Y}
\end{align}
This expression can be written in a form analogous to the previous case,
\begin{align}
  \braket{\hat{Y}_{\hat{\Pi}} } &= \ii|A_{n}|^{2}l_{B}^{2}(2\alpha)^{n-1}\exp\left( \frac{|\zeta|^{2}}{2\alpha}\right) \nonumber \\
  & \quad \times\Big[ \sqrt{2\alpha}\left(\Upsilon_{2} - n\Upsilon_{1} \right) 
  - 2(\ii\zeta_{I}+\alpha\kappa) \Upsilon_{0}\Big]. \label{eq:app_val-esp-upsilons-Y}
\end{align}
Then, by comparing Eqs. (\ref{eq:app_val-esp-upsilons-Y}) and (\ref{eq:app_val-esp-upsilons}), one directly obtains
\begin{equation}
  \braket{\hat{Y}_{\hat{\Pi}} } = 2l_{B}^{2} \zeta_{I}.
\end{equation}
\section{\label{app:Expected-value} Expectation value of the anti-Hermitian component}

The imaginary part of the perturbation that characterizes the system studied in this work, $\hat{U} = \ii e \mathcal{E} y$, constitutes an anti-Hermitian component of the total Hamiltonian. To show that its expectation value is directly proportional to the imaginary part of the parameter $\zeta$, it suffices to evaluate the following expectation value,
\begin{align}
  \braket{\psi_{n}| \hat{y} |\psi_{n}} 
  &= \iint\limits_{\mathbb{R}^{2}}\psi_{n}^{*}(x,y)y\psi_{n}(x,y)dxdy \nonumber \\ 
  &= \iint\limits_{\mathbb{R}^{2}}|\psi_{n}|^{2}  
  y\, dxdy \label{eq:app_valor-esperado}
\end{align}
where 
\begin{align}
  |\psi_{n}|^{2} &= |A_{n}|^{2}(2\alpha)^{2n} \exp\!\big[ -2\alpha (x^{2}+y^{2})+2\zeta_{R}x -2\zeta_{I}y \nonumber \\
  & \quad +2\kappa(2\alpha x -\zeta_{R}) \big] 
  \left[\left(x-\frac{\zeta_{R}}{2\alpha}\right)^{2} 
  + \left(y + \frac{\zeta_{I}}{2\alpha}\right)^{2} \right]^{n}, \label{eq:app_psi-density}
\end{align}
where the definitions in Eqs. (\ref{eq:app_def-k-z}) have been used. Upon performing the change of variables (\ref{eq:app_xix})--(\ref{eq:app_difs-xis}), the following simplified expression for the expectation value is obtained
\begin{equation}
  \braket{\psi_{n}| y | \psi_{n}} = |A_{n}|^{2}(2\alpha)^{n-3/2}\exp\! \left(\frac{|\zeta|^{2}}{2\alpha}\right)
  \left(\Upsilon_{3} - \frac{\zeta_{I}}{\sqrt{2\alpha}} \Upsilon_{0} \right) \label{eq:app_ve-upsilons}
\end{equation}
where $\Upsilon_{0}$ was defined in Eqs.~\eqref{eq:app_upsilon0} and \eqref{eq:app_upsilon0-resuelto} and
\begin{align}
  \Upsilon_{3} &= \int\limits_{0}^{\infty}\int\limits_{0}^{2\pi}\exp(-\xi_{r}^{2}
  +\gamma\xi_{r}\cos\xi_{\theta} )\xi_{r}^{2n+2}\sin\xi_{\theta} d\xi_{\theta}d\xi_{r} \label{eq:app_upsilon1}
\end{align}
with $\gamma = 2\kappa\sqrt{2\alpha}$. 
One can note that $\Upsilon_{3}$ vanishes due to the following integral, 
\begin{equation}
  \int\limits_{0}^{2\pi}\exp(\gamma\xi_{r}\cos\xi_{\theta})\sin\xi_{\theta}d\xi_{\theta} = 0. \label{eq:app_upsilon1-cero}
\end{equation}
Finally, the expectation value of $\hat{U}$ can be written as 
\begin{align}
  \braket{\hat{U}} &= -\ii\zeta_{I}|A_{n}|^{2}(2\alpha)^{n-2}\pi e\mathcal{E}\exp\! \left(\frac{|\zeta|^{2}}{2\alpha}\right) \nonumber \\ 
  & \quad \times \Gamma(n+1)\,{}_{1}F_{1}\left(n+1,1,\frac{\gamma^{2}}{4}\right) \nonumber \\
  &= - \frac{\ii e\mathcal{E}\zeta_{I}}{2\alpha}. \label{eq:app_valor-esperado-resuelto}
\end{align}\\
\section{\label{app:current-vector} Expectation value of the current vector}

The expectation value $\braket{\bf J_{n}}$ is determined by the quantities $\braket{\psi_{n}^{2} | v_{n_{x}} | \psi_{n}^{2}}$ and $\braket{\psi_{n}^{2} | v_{n_{y}} | \psi_{n}^{2}}$, where $v_{n_{x}}$ and $v_{n_{y}}$, are defined in Eqs.~(\ref{eq:current-compX}) and (\ref{eq:current-compY}), respectively.
It is straightforward to verify that the $X$-component of this value vanishes due to the presence of an integral of the form given in Eq.(\ref{eq:app_upsilon1-cero}). The $Y$-component, in turn, requires the evaluation of the following integrals,
\begin{align}
  \braket{\psi_{n}^{2} | v_{n_{y}} | \psi_{n}^{2}} &=
  - \frac{\omega_{c}}{\sqrt{8\alpha}}
  |A_{n}|^{4} (2\alpha)^{2n-1}\exp\left( \frac{|\zeta|^{2}}{\alpha}\right) \nonumber \\
  & \quad \times \left(\frac{\gamma}{2}\Upsilon_{4}+n\Upsilon_{5}+\Upsilon_{6} \right), \label{eq:app_val-esp-vny}
\end{align} 
where
\begin{align}
  \Upsilon_{4} &= \int\limits_{0}^{\infty}\int\limits_{0}^{2\pi}\exp(-2\xi_{r}^{2}+2\gamma\xi_{r}\cos\xi_{\theta})\xi_{r}^{4n+1}d\xi_{\theta}d\xi_{r}, \\
  \Upsilon_{5} &= \int\limits_{0}^{\infty}\int\limits_{0}^{2\pi}\exp(-2\xi_{r}^{2}+2\gamma\xi_{r}\cos\xi_{\theta})\xi_{r}^{4n}\cos\xi_{\theta}d\xi_{\theta}d\xi_{r}, \\
  \Upsilon_{6} &= \int\limits_{0}^{\infty}\int\limits_{0}^{2\pi}\exp(-2\xi_{r}^{2}+2\gamma\xi_{r}\cos\xi_{\theta})\xi_{r}^{4n+2}\cos\xi_{\theta}d\xi_{\theta}d\xi_{r}.
\end{align}
With the aid of Eqs. (\ref{eq:app_clave1}) and (\ref{eq:app_clave2}), one obtains
\begin{align}
  \Upsilon_{4} &= \frac{\pi}{2^{2n+1}} \Gamma(2n+1)
  {}_{1}F_{1}\left(2n+1,1, \frac{\gamma^{2}}{2} \right), \\
  \Upsilon_{5} &= \frac{\pi\gamma}{2^{2n+1}} \Gamma(2n+1)
  {}_{1}F_{1}\left(2n+1,2, \frac{\gamma^{2}}{2} \right), \\
  \Upsilon_{6} &= \frac{\pi\gamma}{2^{2n+2}} \Gamma(2n+2)
  {}_{1}F_{1}\left(2n+2,2, \frac{\gamma^{2}}{2} \right).
\end{align}
Then, using Eq.(\ref{eq:app_clave6}), one finds
\begin{equation}
  \frac{\gamma}{2}\Upsilon_{4}+n\Upsilon_{5} =
  \Upsilon_{6},
\end{equation}
which brings Eq.(\ref{eq:app_val-esp-vny}) to the form
\begin{align}
  \braket{\psi_{n}^{2} | v_{n_{y}} | \psi_{n}^{2}} &=
  - \frac{\omega_{c}}{\sqrt{8\alpha}}
  |A_{n}|^{4} \frac{\pi\gamma}{4}\alpha^{2n-1}\exp\left( \frac{|\zeta|^{2}}{\alpha}\right) \nonumber \\
  & \quad \times (2n+1)!{}_{1}F_{1}\left(2n+2,2, \frac{\gamma^{2}}{2} \right).
\end{align} 
Finally, substituting $|A_{n}|^{4}$ and using the identity (\ref{eq:app_clave4}), one obtains
\begin{align}
  \bra{\psi_{n}^{2}}v_{n_{y}}\ket{\psi_{n}^{2}} &= -
  \frac{(\frac{1}{2})_{n}}{n!}\frac{e\mathcal{E}}{2\pi\hbar}
  \frac{L_{2n}^{(1)}\left(-\frac{\gamma^{2}}{2}\right)}{\left[L_{n}\left(-\frac{\gamma^{2}}{4}\right)\right]^{2}},
\end{align}
where $(a)_{n}$ denotes the Pochhammer symbol, $L_{n}^{(1)}(x)$ is the first associated Laguerre polynomial, and the definitions (\ref{substitution}) and (\ref{eq:gamma}) have been used to express the result in terms of the charge $e$ and electric field intensity $\mathcal{E}$.\\

\section{\label{app:NEGF}NEGF details}

In this appendix, we summarize the NEGF procedure used to calculate the resistance. Detailed discussions can be found in Refs.~\cite{Datta2005,Datta1997,DiVentra2008}, and a concise introduction is provided in Ref.~\cite{stegmann2021brief}. The tight-binding Hamiltonian is given by Eq.~\eqref{TBH2}. We consider transport without spin degeneracy in the low-temperature, stationary linear-response regime.

The non-Hermitian contribution is treated phenomenologically. We assume that the environmental response compensates particle exchange associated with gain and loss so that the terminal currents admit a current-conserving Landauer--B\"uttiker description, with no additional injection term beyond the terminals included below. This is an assumption of the transport model.

The retarded Green's function includes the source, drain, and attached probes:
\begin{equation}
  \label{eq:9}
  G(E)=
  \left(
  E-H-\Sigma_S-\Sigma_D-\sum_{p\in P}\Sigma_p
  \right)^{-1},
\end{equation}
where $E$ is the energy at which the transport response is evaluated and $P$ denotes the set of probes. The parameters are assumed to yield a stable stationary response. The self-energy of each terminal is modeled within the wide-band approximation,
\begin{equation}
  \label{eq:12b}
  \Sigma_k=
  -\ii\eta\sum_{i\in\mathrm{contact}\ k}\ket{i}\bra{i},
\end{equation}
where $\eta>0$ describes the contact coupling and $k$ runs over the source, drain, and probes. The corresponding broadening matrices are
\begin{equation}
  \Gamma_k=\ii(\Sigma_k-\Sigma_k^\dagger)
  =-2\operatorname{Im}(\Sigma_k).
\end{equation}
The transmission between source and drain is
\begin{equation}
  T_{SD}(E)=
  \operatorname{Tr}\!\left(
  \Gamma_SG\Gamma_DG^\dagger
  \right).
\end{equation}
Throughout this appendix, the first index of a transmission labels the receiving terminal. Thus, $T_{pk}$ describes transmission from terminal $k$ into terminal $p$.

For the assumed current-conserving terminal response, we impose
\begin{equation}
  \sum_k T_{pk}(E)=\sum_k T_{kp}(E),
\end{equation}
where the sums include all terminals. This balance condition is required for the probe elimination used below.

The probes absorb and incoherently reinject electrons. Their occupations are determined by requiring zero net current at each probe at the energy under consideration. Eliminating these occupations gives the effective transmission
\begin{equation}
  \widetilde{T}_{SD}(E)
  =
  T_{SD}(E)
  +\mathbf{T}_{S,P}\,
  \mathbf{M}^{-1}\,
  \mathbf{T}_{P,D},
\end{equation}
where $\mathbf{T}_{S,P}$ contains the transmissions from the probes into the source, $\mathbf{T}_{P,D}$ contains those from the drain into the probes, and
\begin{equation}
  \mathbf{M}_{pp'}
  =
  \delta_{pp'}\sum_k T_{pk}-T_{pp'}
\end{equation}
is the probe conservation matrix. The participating probes are assumed to be connected to the transport network so that $\mathbf{M}$ is invertible. Without spin degeneracy, the resulting effective conductance is
$\mathcal{G}=(e^2/h)\widetilde{T}_{SD}$.

To identify the corresponding resistance with the Hall resistance, we additionally assume that, within the plateau interval considered, the two transverse voltage probes sample edge branches whose electrochemical potentials equal those of the source and drain, respectively. Their voltage difference therefore equals the source--drain bias. Under this measurement condition, the Hall resistance coincides with the effective source--drain resistance:
\begin{equation}
  \label{eq:R_effective}
  R(E)=\frac{h}{e^2}\frac{1}{\widetilde{T}_{SD}(E)}.
\end{equation}

We also calculate the trace of the spectral function,
\begin{equation}
  \hat{A}(E)=
  \ii\operatorname{Tr}\!\left[G(E)-G^\dagger(E)\right].
\end{equation}
The corresponding density-of-states normalization is
$\hat{A}(E)/(2\pi)$.

\bibliographystyle{unsrt}
\bibliography{bib}

\end{document}